%% file: arxiv.tex
\documentclass[conference]{IEEEtran}
\makeatletter
\def\ps@headings{%
\def\@oddhead{\mbox{}\scriptsize\rightmark \hfil \thepage}%
\def\@evenhead{\scriptsize\thepage \hfil \leftmark\mbox{}}%
\def\@oddfoot{}%
\def\@evenfoot{}}
\makeatother
\usepackage{amsmath,amssymb}
\newtheorem{theorem}{Theorem}

\usepackage{graphicx,tikz,scalefnt}
\usepackage[ruled,linesnumbered,nofillcomment,noline]{algorithm2e}
\SetCommentSty{textrm}
\title{Use of Devolved Controllers\\ in Data Center Networks}

\author{
\IEEEauthorblockN{Adrian S.-W. Tam \hspace{2em}
Kang Xi \hspace{2em}
H. Jonathan Chao}\\
\IEEEauthorblockA{Department of Electrical and Computer Engineering\\
Polytechnic Institute of
New York University\\
Email: adrian@antioch.poly.edu, kxi@poly.edu, chao@poly.edu}
}

\begin{document}

\maketitle

\begin{abstract}
In a data center network, for example, it is quite often to use controllers to
manage resources in a centralized manner. Centralized control, however, imposes
a scalability problem. In this paper, we investigate the use of multiple
independent controllers instead of a single omniscient controller to manage
resources.  Each controller looks after a portion of the network only, but they
together cover the whole network.  This therefore solves the scalability
problem. We use flow allocation as an example to see how this approach can
manage the bandwidth use in a distributed manner. The focus is on how to assign
components of a network to the controllers so that (1) each controller only
need to look after a small part of the network but (2) there is at least one
controller that can answer any request. We outline a way to configure the
controllers to fulfill these requirements as a proof that the use of devolved
controllers is possible. We also discuss several issues related to such
implementation.
\end{abstract}

\section{Introduction}\label{sec:intro}
Among recent years' literature on data center networking, using a centralized
controller for coordination or resource management is a common practice
\cite{arrhv10,cdghwbcfg06,cfpms07,dg04,ggl03,gjkklmps09,mpfhmrsv09}. In
\cite{ggl03}, for example, a master server is used to hold the metadata for a
distributed file system. In another example, \cite{arrhv10}, a flow scheduling
server is responsible for computing a new route for a rerouted flow at real
time. In \cite{cfpms07}, a controller is also used to enforce a route for a
packet so that its use of the network compliances with the policies. Using a
centralized controller not only makes the design simpler, but also sufficient.
In \cite{ggl03}, the authors claim that a single controller is enough to drive
a fairly large network and the problem of single point of failure can be
mitigated by replication.

Nevertheless, the use of a centralized controller subjects to scalability
constraints. Usually the scalability problems are solved by load balancing. For
example, replicating the whole database to multiple servers is a common way to
load balance MySQL servers \cite{s08}. However, if the scalability problem is
caused by too much data stored in a controller so that its response time is
degraded, balancing load by identical controllers cannot solve the problem. As
the data center network grows larger and larger, we can expect to have such
problem in the near future. Therefore, it is interesting to study an
alternative solution to a \emph{single} centralized controller.

We study the use of \emph{devolved} controllers in this paper. They together
function as a single logical centralized controller but none of them have the
complete information of the whole data center network. This is beneficial, for
example, when the controllers are supposed to provide real time computations
and too much data would cause the computation slow.

We take the following flow route assignment as an example to see how we can use
devolved controllers: Whenever a flow, identified by a source and a destination
node in the network, is to be established, the sending node will query the
controllers for the route it should use to avoid congestion. The controllers
are therefore responsible to monitor the network to assist the route selection.
If the network topology were too large, the response time would be too long to
be useful. Thus instead of a single omniscient controller to cover the whole
network, we use multiple `smaller' controllers so that each of them covers a
partial topology only. When a controller is asked for a route, it responds with
the topology data it has.

Note that this paper is not about route optimality or routing protocols, but to
show that an omniscient controller is not the only solution. The novelity of
this paper is on the concept of \emph{devolved controllers}, which eliminates
the scalability problem of traditional omniscient controller.

Our work is on the control aspect of data center networks. In recent years,
there are many literatures that focus on control plane design in networks.
Examples are OpenFlow \cite{mabpprst08}, NOX \cite{gkppcms08} and Ethane
\cite{cfpms07}. To address the scalability issue of the controllers in these
designs, their developers proposed \cite{m10} to partition the controllers
horizontally (i.e. replication of controllers) and vertically (i.e. each
controller serve a part of the network). While horizontal partitioning is
trivial, this paper explore into the ways of vertical partitioning.

In the rest of this paper, we describe an example on controller use in section
\ref{sec:formulation} and provide heuristic algorithms in section
\ref{sec:algo} on how to configure the controllers. Evaluation is provided in
section \ref{sec:eval} and discussion on the use of devolved controllers in
section \ref{sec:discussion}.

\section{Problem statement}\label{sec:formulation}

On a network represented by a connected graph $G=(V,E)$, a flow is identified
by the ordered pair $(s,t)$ where $s,t\in V$. On such a network, there are $q$
controllers. Each of them is managing a portion of the network, represented by
a subgraph of $G$. We say a controller that manages $G'=(V',E')$ \emph{covers}
a node $v\in V$ or a link $e\in E$ if $v\in V'$ or $e\in E'$, respectively.
Upon a flow is going to be established, the source node $s$ queries the
controllers for a route to destination $t$. Among the $q$ controllers, at least
one of them responds with a path $p$ that connects $s$ to $t$, which is the
route for this flow. \tablename~\ref{tab:terms} summarizes the terms used in
this paper.

The controllers are supposed to respond to the flow route query in a very short
time. Therefore, computationally intensive path-finding algorithms are not
viable. Furthermore, we have to ensure at least one the $q$ controllers
can provide a route for any source-destination pair $(s,t)$. One way is
to have all routes pre-computed. Assume for any ordered pair $(s,t)$, we
compute $k$ different paths $p_1,\ldots,p_k$ that join $s$ to $t$. We call the
set $M=\{p_1,\ldots,p_k\}$ as a $k$-multipath. Then, we install the multipath
into a controller. Upon the query is issued, the controllers will return the
least congested one of the $k$ paths.

\begin{figure}
  \centering
  \includegraphics[scale=0.6]{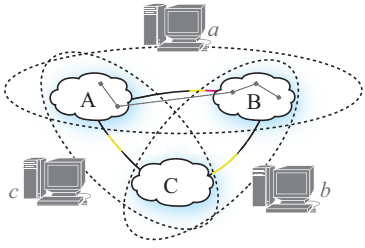}
  \caption{A network managed by three controllers}
  \label{fig:cloud}
\end{figure}

\figurename~\ref{fig:cloud} gives an example of a network with three
controllers. The part of the network that a controller covers is illustrated by
a dotted ellipse. Precisely, there is a controller that covers all the routes
between nodes in regions A and B as well as the routes within those regions;
another controller covers that of regions A and C and yet another is for
regions B and C. In this way, none of the controllers monitor every
spot in the network but they together can respond to any request from any node.
For instance, the route illustrated in \figurename~\ref{fig:cloud} is entirely
within the jurisdiction of controller $a$. So $a$ is the one to provide this
route upon request. When the network grows, we can install more controllers to
cover the network so that none of the controllers need to manage a region of
too large in size.

Assume we have all the paths pre-computed, the immediate questions are then
\begin{enumerate}
\item How to optimally allocate the multipaths into the $q$ controllers? We
define the optimality as the smallest number of unique links to be monitored by
the controller. In other words, we prefer the controller to cover as small
portion of the network as possible.
\item If no controller can monitor more than $N$ links on the network, what is
the number of controllers needed?
\end{enumerate}

\section{Approximate solution}\label{sec:algo}

We first consider the problem of optimal allocation of the multipaths into $q$
controllers.

\begin{table}
   \centering
   \begin{tabular}{ll}
      $(s,t)$ & Source-destination pair \\
      flow & A data stream from a source node to destination node \\
      path/route& A path that connects two nodes \\
      multipath & A set of paths that each of them connects the same pair \\
%      route & A path that is selected for a flow to use \\
   \end{tabular}
   \caption{Definition of terms used in this paper}
   \label{tab:terms}
\end{table}

For $n=|V|$ nodes on the network, there are $n(n-1)$ different
source-destination pairs. This is also the number of multipaths to be
pre-computed as mentioned in section \ref{sec:formulation}. It is a NP-hard
problem to find the optimal allocation to $q$ controllers\footnote{The problem
in concern is an extended problem of graph partitioning. It is well-known in
algorithmic graph theory that graph partitioning is NP-hard. Using heuristic
algorithms such as Kernighan-Lin \cite{kl70} is the standard way to solve graph
partitioning problems.}. The size of the solution space for allocating $n(n-1)$
multipaths to $q$ controllers is given by the Stirling number of the second
kind \cite{jk77}:
$$ \frac{1}{q!}\sum_{j=0}^{q} (-1)^j {q \choose j} (q-j)^{n(n-1)} $$
It becomes
intractable quickly for a moderately large network.

\subsection{Path-partition approach}\label{sub:pathpart}

Instead of looking for a global optimal solution, we developed a heuristic
algorithm to obtain an approximate solution. The algorithm is in two parts:
Firstly it enumerates all the $k$-multipaths for all source-destination pairs
$(s,t)$. Then, it allocates each multipath into one of the controllers
according to a cost function. The multipaths are pre-computed with no knowledge
of where they are to be allocated to the controllers. We call this the
\emph{path-partition approach}:

%\begin{algorithm}
%\caption{Path-partition heuristic algorithm}\label{alg:pathpart}
%\KwData{Network $G=(V,E)$, $q=$number of controllers}
%\tcc{Enumerating multipaths}
%$\mathcal{M} := \emptyset$\;
%\ForEach{$s,t\in V$}{\label{alg:mpathenum1}
%  Set link weight $w(e)=1$ for all links $e\in E$\;
%  $M := \emptyset$\;
%  \For{$i:=1$ \KwTo $k$}{
%    Find a path $p$ joining $s$ to $t$ using Dijkstra's algorithm with link weight $w(e)$\;\label{alg:dijkstra1}
%    $M := M \cup \{p\}$\;
%    \ForEach{link $e$ on path $p$}{
%      $w(e) := w(e) + \omega$ \tcc*[l]{Update link weight}\label{alg:linkweight}
%    }
%  }
%  $\mathcal{M} := \mathcal{M} \cup \{M\}$\;
%}\label{alg:mpathenum2}
%\tcc{Partition multipath into $q$ controllers}
%\ForEach{$M\in\mathcal{M}$ in random order}{\label{alg:partition1}
%  \For{$i:=1$ \KwTo $q$}{
%    $c_i :=$ cost of adding multipath $M$ to controller $i$\label{alg:cost}
%  }
%  Allocate multipath $M$ to the controller $j = \arg\min c_j$
%}\label{alg:partition2}
%\end{algorithm}

\begin{algorithm}
\scalefont{0.9}
\caption{Path-partition heuristic algorithm}\label{alg:pathpart}
\KwData{Network $G=(V,E)$,
$q=$ number of controllers}
\ForEach{$s,t\in V$ in random order}{\label{alg:mpathenum1}
  \tcc{Constructing a $k$-multipath from $s$ to $t$}
  $M := k$ paths joining $s$ to $t$\;\label{alg:pathfind1}
  \tcc{Allocate into a controller}
  \For{$i:=1$ \KwTo $q$}{\label{alg:partition1}
    $c_i :=$ cost of adding multipath $M$ to controller $i$\label{alg:cost}
  }
  Allocate multipath $M$ to the controller $j = \arg\min c_j$\label{alg:partition2}
}
\end{algorithm}

We implemented the multipath enumeration in algorithm \ref{alg:pathpart} (line
\ref{alg:pathfind1}) after \cite{myam09}. We find a path from $s$ to $t$ using
Dijkstra's algorithm with unit link weight for each link in $E$.  Then, the
links used in this path have their link weights increased by an amount
$\omega$. The Dijkstra's algorithm and link weight modification are repeated
until all $k$ paths are found. It is a straightforward algorithm to find $k$
distinct paths from $s$ to $t$ by using Dijkstra's algorithm iteratively with
modified link weight. The link weight increase is suggested in \cite{myam09} to
be $\omega=|E|$ to prefer as much link-disjointness as possible between paths.
When a short route is preferred, however, $\omega$ should set to a small value.
We use the latter approach.

Other methods to compute multipaths in algorithm \ref{alg:pathpart} are
available, such as \cite{m78} or \cite{go02}. The way the multipaths are
found does not affect the discussion hereinafter.

The essential part of the heuristic algorithm is the lines
\ref{alg:partition1}--\ref{alg:partition2}. It allocates the multipaths to
controllers one by one according to a cost function. The goal of the cost
function is to allocate the multipath into controllers such that the maximum
number of links to be monitored by a controller is minimized. With that in
mind, we established the following heuristic:
\begin{enumerate}
\item A multipath shall be allocated to a controller if that
controller already monitors most of the links used by that multipath; and
\item In an optimal allocation, the total number of links
monitored by each controller shall be roughly the same.
\end{enumerate}
Therefore, we define the cost function in line \ref{alg:cost} as
$$ c_i = \alpha \nu_i(M) + \mu_i $$
where $\mu_i$ is the number of links already monitored by controller $i$ at the
moment and $\nu_i(M)$ is the number of links in the multipath $M$ that is not
yet monitored by that controller, i.e. if $M$ is allocated to
controller $i$, the total number of links monitored by controller $i$ would
become $\mu_i + \nu_i(M)$. Parameter $\alpha$ adjusts their weight in the cost
function. When $\alpha \approx 0$, we ignore the benefit of reusing existing
links in a controller. When $\alpha \approx \infty$, however, we do not require
the controllers to be balanced. This usually yields the result that almost all
multipaths are allocated to the same controller, which can be explained by the
Matthew's effect on the allocation process of lines
\ref{alg:partition1}--\ref{alg:partition2}.  We empirically found that $\alpha$
between 4 to 8 gives a good result. We set $\alpha=4$ in our experiments but
the wide range of appropriate values for $\alpha$ suggests that it is not very
sensitive.

\subsection{Partition-path approach}

Another way to allocate multipaths into controllers, the \emph{partition-path
approach}, is available. Its idea is that, if a controller is already
monitoring certain links, we can find the $k$-multipath between the
source-destination pair $(s,t)$ that uses those links as long as it is
possible. Therefore, in this approach, we first partition the links into
$q$ controllers as their preferred links. Then the multipath connecting $s$ to
$t$ is computed individually in each controller, with preference given to
certain links. Algorithm \ref{alg:partpath} illustrates the idea.

The algorithm begins with a procedure to randomly partition the set of links
$E$ into the $q$ controllers such that each controller $i$ covers a subset
$\mathcal{E}_i\subset E$ preliminarily (lines
\ref{alg:part3}--\ref{alg:part4}). Then for each source-destination pair
$(s,t)$, it finds a $k$-multipath on each controller $i$ with the preference to
using links in $\mathcal{E}_i$. The same path-finding algorithm is used in
Algorithm \ref{alg:partpath} as in Algorithm \ref{alg:pathpart}. Note that, the
links in $\mathcal{E}_i$ affects the path-finding algorithm by changing the
initial link weight only. In the other part of the algorithm, such as the cost
function in line \ref{alg:cost2}, it is not involved. The same cost function as
in section \ref{sub:pathpart} is used here.

%\begin{algorithm}
%\caption{Partition-path heuristic algorithm}\label{alg:partpath}
%\KwData{Network $G=(V,E)$, $q=$number of controllers}
%\tcc{Partition links to controllers preliminarily}
%\lForEach{$i:=1$ \KwTo $q$}{$\mathcal{E}_i = \emptyset$}\;\label{alg:part3}
%\ForEach{$e\in E$}{\label{alg:partlink}
%   $i:=$ random integer in $\{1,\ldots,q\}$\;
%   $\mathcal{E}_i := \mathcal{E}_i \cup \{e\}$
%}\label{alg:part4}
%\tcc{Enumerating multipaths and allocating them into controllers}
%\ForEach{$s,t\in V$ in random order}{
%  \ForEach{$i:=1$ \KwTo $q$}{
%    Set link weight $w(e)=1$ for all $e\in \mathcal{E}_i$\;
%    Set link weight $w(e)=\psi$ for all other $e\in E$\;
%    $M_i := \emptyset$\;
%    \For(\tcc*[f]{Finds $k$-multipath}){$j:=1$ \KwTo $k$}{
%      Find a path $p$ joining $s$ to $t$ using Dijkstra's algorithm with link weight $w(e)$\;\label{alg:dijkstra}
%      $M_i := M_i \cup \{p\}$\;
%      \ForEach{link $e$ on path $p$}{
%        $w(e) := w(e) + \omega$
%      }
%    }
%    $c_i :=$ cost of adding multipath $M_i$ to controller $i$\label{alg:cost2}
%  }
%  Allocate multipath $M_j$ to controller $j = \arg\min c_j$
%  $\mathcal{E}_j := \mathcal{E}_j \cup \{e: \textrm{for all links $e$ in path }p\}$
%}
%\end{algorithm}

\begin{algorithm}
\scalefont{0.9}
\caption{Partition-path heuristic algorithm}\label{alg:partpath}
\DontPrintSemicolon
\KwData{Network $G=(V,E)$, $q=$ number of controllers}
\tcc{Partition links to controllers preliminarily}
\lForEach{$i:=1$ \KwTo $q$}{$\mathcal{E}_i = \emptyset$}\;\label{alg:part3}
\ForEach{$e\in E$}{\label{alg:partlink}
   $i:=$ random integer in $\{1,\ldots,q\}$\;
   $\mathcal{E}_i := \mathcal{E}_i \cup \{e\}$
}\label{alg:part4}
\tcc{Enumerate multipaths and allocate into controllers}
\ForEach{$s,t\in V$ in random order}{
  \ForEach{$i:=1$ \KwTo $q$}{
    Set link weight $w(e)=\begin{cases} 1 & \textrm{for all $e\in \mathcal{E}_i$} \\
        \psi & \textrm{for all other $e\in E$}\end{cases}$\;
    $M_i := k$ paths joining $s$ to $t$\;\label{alg:dijkstra}
    $c_i :=$ cost of adding multipath $M_i$ to controller $i$\label{alg:cost2}
  }
  Allocate multipath $M_j$ to controller $j = \arg\min c_j$\;\label{alg:partition}
  $\mathcal{E}_j := \mathcal{E}_j \cup \{e: \textrm{for all links $e$ in $M_j$}\}$
}
\end{algorithm}

\section{Performance}\label{sec:eval}

We applied the heuristic algorithms on different topologies from the Rocketfuel
project \cite{smw02} to evaluate its performance. We use Rocketfuel topologies
as there is no detailed data center topologies available publicly. We also
evaluate the algorithm with a fat tree topology, which is likely to be used in
data center networks, in section \ref{sub:regular}.

\subsection{Size of controllers}\label{sub:size}

\begin{figure}
\centering
{\scalefont{0.5}
\begin{tikzpicture}[every node/.style={circle,draw},scale=0.14]
\node (14) at (0.687765,17.7997109) {14};
\node (26) at (1.63931433,10.2674497) {26};
\node (6) at (30.6161235,12.090556) {6};
\node (10) at (3.48259507,24.8586487) {10};
\node (3) at (30.6161235,19.6824235) {3};
\node (13) at (16.7899308,31.1204845) {13};
\node (15) at (29.2077924,23.2401084) {15};
\node (8) at (12.9712479,0.89275424) {8};
\node (4) at (9.332442,2.07511045) {4};
\node (7) at (20.548302,1.36946356) {7};
\node (25) at (32.8169715,23.0958541) {25};
\node (11) at (9.332442,29.6980454) {11};
\node (12) at (26.9586245,26.3354036) {12};
\node (27) at (1.63931433,21.5055298) {27};
\node (0) at (29.2077924,8.5328711) {0};
\node (17) at (6.10171,4.1251792) {17};
\node (18) at (3.48259507,6.9143308) {18};
\node (2) at (20.548302,30.4034454) {2};
\node (16) at (30.9913963,26.4055909) {16};
\node (23) at (24.0104052,2.99858486) {23};
\node (1) at (24.0104052,28.7743241) {1};
\node (9) at (26.9586245,5.4375759) {9};
\node (24) at (16.7899308,0.652495) {24};
\node (5) at (31.0957955,15.8866661) {5};
\node (21) at (34.2588091,15.8866661) {21};
\node (19) at (6.10171,27.6478003) {19};
\node (20) at (12.9712479,30.8802958) {20};
\node (22) at (0.687765,13.9732686) {22};
\draw (14) -- (26);
\draw (14) -- (6);
\draw (14) -- (10);
\draw (3) -- (13);
\draw (3) -- (15);
\draw (3) -- (8);
\draw (3) -- (4);
\draw (3) -- (7);
\draw (15) -- (25);
\draw (4) -- (8);
\draw (4) -- (26);
\draw (15) -- (6);
\draw (11) -- (10);
\draw (12) -- (13);
\draw (12) -- (15);
\draw (14) -- (27);
\draw (6) -- (0);
\draw (27) -- (10);
\draw (17) -- (18);
\draw (18) -- (2);
\draw (15) -- (16);
\draw (4) -- (23);
\draw (17) -- (1);
\draw (17) -- (7);
\draw (17) -- (8);
\draw (17) -- (14);
\draw (17) -- (15);
\draw (7) -- (8);
\draw (7) -- (23);
\draw (7) -- (9);
\draw (7) -- (24);
\draw (7) -- (26);
\draw (5) -- (0);
\draw (5) -- (6);
\draw (5) -- (21);
\draw (5) -- (3);
\draw (19) -- (10);
\draw (19) -- (20);
\draw (8) -- (24);
\draw (23) -- (9);
\draw (23) -- (0);
\draw (9) -- (24);
\draw (13) -- (1);
\draw (13) -- (14);
\draw (13) -- (22);
\draw (13) -- (9);
\draw (13) -- (20);
\draw (13) -- (15);
\draw (9) -- (0);
\draw (9) -- (6);
\draw (13) -- (2);
\draw (13) -- (8);
\draw (13) -- (26);
\draw (20) -- (11);
\draw (0) -- (14);
\draw (0) -- (3);
\draw (20) -- (5);
\draw (20) -- (6);
\draw (6) -- (7);
\draw (7) -- (10);
\draw (8) -- (23);
\draw (22) -- (14);
\draw (22) -- (8);
\draw (8) -- (9);
\draw (2) -- (1);
\draw (1) -- (5);
\end{tikzpicture}}
\caption{Topology of an irregular network with 28 nodes and 66 links.}\label{fig:ebone}
\end{figure}
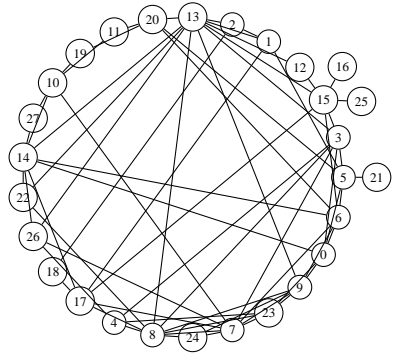

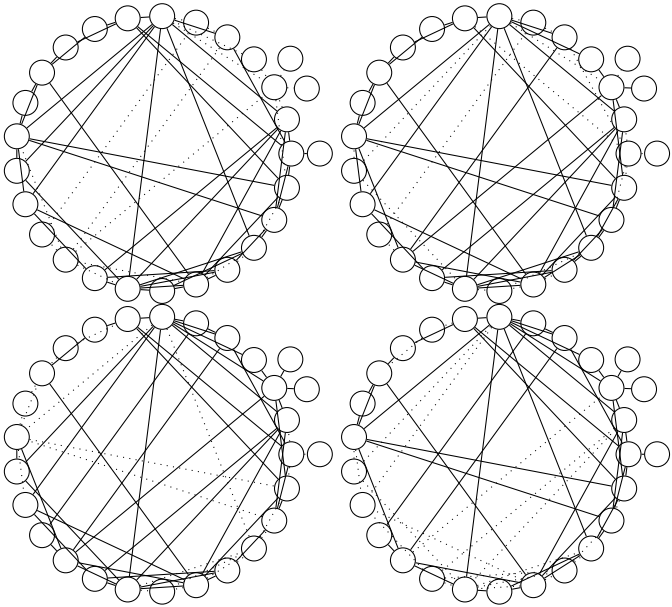
\begin{figure}
\begin{center}
\begin{tikzpicture}[every node/.style={circle,draw},scale=0.24]
\node (14) at (0.3438825,8.89985545) {};
\node (26) at (0.819657165,5.13372485) {};
\node (6) at (15.30806175,6.045278) {};
\node (10) at (1.741297535,12.42932435) {};
\node (3) at (15.30806175,9.84121175) {};
\node (13) at (8.3949654,15.56024225) {};
\node (15) at (14.6038962,11.6200542) {};
\node (8) at (6.48562395,0.44637712) {};
\node (4) at (4.666221,1.037555225) {};
\node (7) at (10.274151,0.68473178) {};
\node (25) at (16.40848575,11.54792705) {};
\node (11) at (4.666221,14.8490227) {};
\node (12) at (13.47931225,13.1677018) {};
\node (27) at (0.819657165,10.7527649) {};
\node (0) at (14.6038962,4.26643555) {};
\node (17) at (3.050855,2.0625896) {};
\node (18) at (1.741297535,3.4571654) {};
\node (2) at (10.274151,15.2017227) {};
\node (16) at (15.49569815,13.20279545) {};
\node (23) at (12.0052026,1.49929243) {};
\node (1) at (12.0052026,14.38716205) {};
\node (9) at (13.47931225,2.71878795) {};
\node (24) at (8.3949654,0.3262475) {};
\node (5) at (15.54789775,7.94333305) {};
\node (21) at (17.12940455,7.94333305) {};
\node (19) at (3.050855,13.82390015) {};
\node (20) at (6.48562395,15.4401479) {};
\node (22) at (0.3438825,6.9866343) {};
\draw[black] (14) -- (26);
\draw[black] (14) -- (6);
\draw[black] (14) -- (10);
\draw[black] (3) -- (13);
\draw[dotted] (3) -- (15);
\draw[black] (3) -- (8);
\draw[black] (3) -- (4);
\draw[black] (3) -- (7);
\draw[dotted] (15) -- (25);
\draw[black] (4) -- (8);
\draw[black] (4) -- (26);
\draw[dotted] (15) -- (6);
\draw[black] (11) -- (10);
\draw[dotted] (12) -- (13);
\draw[dotted] (12) -- (15);
\draw[black] (14) -- (27);
\draw[black] (6) -- (0);
\draw[black] (27) -- (10);
\draw[dotted] (17) -- (18);
\draw[dotted] (18) -- (2);
\draw[dotted] (15) -- (16);
\draw[black] (4) -- (23);
\draw[dotted] (17) -- (1);
\draw[dotted] (17) -- (7);
\draw[dotted] (17) -- (8);
\draw[dotted] (17) -- (14);
\draw[dotted] (17) -- (15);
\draw[black] (7) -- (8);
\draw[black] (7) -- (23);
\draw[dotted] (7) -- (9);
\draw[black] (7) -- (24);
\draw[black] (7) -- (26);
\draw[black] (5) -- (0);
\draw[black] (5) -- (6);
\draw[black] (5) -- (21);
\draw[black] (5) -- (3);
\draw[black] (19) -- (10);
\draw[black] (19) -- (20);
\draw[black] (8) -- (24);
\draw[dotted] (23) -- (9);
\draw[black] (23) -- (0);
\draw[black] (9) -- (24);
\draw[black] (13) -- (1);
\draw[black] (13) -- (14);
\draw[black] (13) -- (22);
\draw[black] (13) -- (9);
\draw[black] (13) -- (20);
\draw[dotted] (13) -- (15);
\draw[black] (9) -- (0);
\draw[dotted] (9) -- (6);
\draw[dotted] (13) -- (2);
\draw[black] (13) -- (8);
\draw[black] (13) -- (26);
\draw[black] (20) -- (11);
\draw[black] (0) -- (14);
\draw[black] (0) -- (3);
\draw[black] (20) -- (5);
\draw[black] (20) -- (6);
\draw[black] (6) -- (7);
\draw[black] (7) -- (10);
\draw[black] (8) -- (23);
\draw[black] (22) -- (14);
\draw[black] (22) -- (8);
\draw[black] (8) -- (9);
\draw[dotted] (2) -- (1);
\draw[black] (1) -- (5);
\end{tikzpicture}
\begin{tikzpicture}[every node/.style={circle,draw},scale=0.24]
\node (14) at (0.3438825,8.89985545) {};
\node (26) at (0.819657165,5.13372485) {};
\node (6) at (15.30806175,6.045278) {};
\node (10) at (1.741297535,12.42932435) {};
\node (3) at (15.30806175,9.84121175) {};
\node (13) at (8.3949654,15.56024225) {};
\node (15) at (14.6038962,11.6200542) {};
\node (8) at (6.48562395,0.44637712) {};
\node (4) at (4.666221,1.037555225) {};
\node (7) at (10.274151,0.68473178) {};
\node (25) at (16.40848575,11.54792705) {};
\node (11) at (4.666221,14.8490227) {};
\node (12) at (13.47931225,13.1677018) {};
\node (27) at (0.819657165,10.7527649) {};
\node (0) at (14.6038962,4.26643555) {};
\node (17) at (3.050855,2.0625896) {};
\node (18) at (1.741297535,3.4571654) {};
\node (2) at (10.274151,15.2017227) {};
\node (16) at (15.49569815,13.20279545) {};
\node (23) at (12.0052026,1.49929243) {};
\node (1) at (12.0052026,14.38716205) {};
\node (9) at (13.47931225,2.71878795) {};
\node (24) at (8.3949654,0.3262475) {};
\node (5) at (15.54789775,7.94333305) {};
\node (21) at (17.12940455,7.94333305) {};
\node (19) at (3.050855,13.82390015) {};
\node (20) at (6.48562395,15.4401479) {};
\node (22) at (0.3438825,6.9866343) {};
\draw[black] (14) -- (26);
\draw[black] (14) -- (6);
\draw[black] (14) -- (10);
\draw[black] (3) -- (13);
\draw[black] (3) -- (15);
\draw[black] (3) -- (8);
\draw[black] (3) -- (4);
\draw[black] (3) -- (7);
\draw[black] (15) -- (25);
\draw[black] (4) -- (8);
\draw[black] (4) -- (26);
\draw[black] (15) -- (6);
\draw[black] (11) -- (10);
\draw[black] (12) -- (13);
\draw[black] (12) -- (15);
\draw[black] (14) -- (27);
\draw[black] (6) -- (0);
\draw[black] (27) -- (10);
\draw[dotted] (17) -- (18);
\draw[dotted] (18) -- (2);
\draw[dotted] (15) -- (16);
\draw[black] (4) -- (23);
\draw[black] (17) -- (1);
\draw[black] (17) -- (7);
\draw[black] (17) -- (8);
\draw[black] (17) -- (14);
\draw[black] (17) -- (15);
\draw[black] (7) -- (8);
\draw[dotted] (7) -- (23);
\draw[black] (7) -- (9);
\draw[dotted] (7) -- (24);
\draw[black] (7) -- (26);
\draw[dotted] (5) -- (0);
\draw[dotted] (5) -- (6);
\draw[dotted] (5) -- (21);
\draw[dotted] (5) -- (3);
\draw[black] (19) -- (10);
\draw[black] (19) -- (20);
\draw[dotted] (8) -- (24);
\draw[black] (23) -- (9);
\draw[black] (23) -- (0);
\draw[dotted] (9) -- (24);
\draw[black] (13) -- (1);
\draw[black] (13) -- (14);
\draw[dotted] (13) -- (22);
\draw[black] (13) -- (9);
\draw[black] (13) -- (20);
\draw[dotted] (13) -- (15);
\draw[black] (9) -- (0);
\draw[black] (9) -- (6);
\draw[dotted] (13) -- (2);
\draw[black] (13) -- (8);
\draw[black] (13) -- (26);
\draw[black] (20) -- (11);
\draw[black] (0) -- (14);
\draw[black] (0) -- (3);
\draw[dotted] (20) -- (5);
\draw[black] (20) -- (6);
\draw[black] (6) -- (7);
\draw[black] (7) -- (10);
\draw[black] (8) -- (23);
\draw[dotted] (22) -- (14);
\draw[dotted] (22) -- (8);
\draw[black] (8) -- (9);
\draw[dotted] (2) -- (1);
\draw[dotted] (1) -- (5);
\end{tikzpicture}\\
\begin{tikzpicture}[every node/.style={circle,draw},scale=0.24]
\node (14) at (0.3438825,8.89985545) {};
\node (26) at (0.819657165,5.13372485) {};
\node (6) at (15.30806175,6.045278) {};
\node (10) at (1.741297535,12.42932435) {};
\node (3) at (15.30806175,9.84121175) {};
\node (13) at (8.3949654,15.56024225) {};
\node (15) at (14.6038962,11.6200542) {};
\node (8) at (6.48562395,0.44637712) {};
\node (4) at (4.666221,1.037555225) {};
\node (7) at (10.274151,0.68473178) {};
\node (25) at (16.40848575,11.54792705) {};
\node (11) at (4.666221,14.8490227) {};
\node (12) at (13.47931225,13.1677018) {};
\node (27) at (0.819657165,10.7527649) {};
\node (0) at (14.6038962,4.26643555) {};
\node (17) at (3.050855,2.0625896) {};
\node (18) at (1.741297535,3.4571654) {};
\node (2) at (10.274151,15.2017227) {};
\node (16) at (15.49569815,13.20279545) {};
\node (23) at (12.0052026,1.49929243) {};
\node (1) at (12.0052026,14.38716205) {};
\node (9) at (13.47931225,2.71878795) {};
\node (24) at (8.3949654,0.3262475) {};
\node (5) at (15.54789775,7.94333305) {};
\node (21) at (17.12940455,7.94333305) {};
\node (19) at (3.050855,13.82390015) {};
\node (20) at (6.48562395,15.4401479) {};
\node (22) at (0.3438825,6.9866343) {};
\draw[dotted] (14) -- (26);
\draw[dotted] (14) -- (6);
\draw[dotted] (14) -- (10);
\draw[black] (3) -- (13);
\draw[black] (3) -- (15);
\draw[black] (3) -- (8);
\draw[black] (3) -- (4);
\draw[black] (3) -- (7);
\draw[black] (15) -- (25);
\draw[black] (4) -- (8);
\draw[black] (4) -- (26);
\draw[black] (15) -- (6);
\draw[black] (11) -- (10);
\draw[black] (12) -- (13);
\draw[black] (12) -- (15);
\draw[dotted] (14) -- (27);
\draw[black] (6) -- (0);
\draw[dotted] (27) -- (10);
\draw[black] (17) -- (18);
\draw[black] (18) -- (2);
\draw[black] (15) -- (16);
\draw[black] (4) -- (23);
\draw[black] (17) -- (1);
\draw[black] (17) -- (7);
\draw[black] (17) -- (8);
\draw[black] (17) -- (14);
\draw[black] (17) -- (15);
\draw[black] (7) -- (8);
\draw[black] (7) -- (23);
\draw[dotted] (7) -- (9);
\draw[dotted] (7) -- (24);
\draw[black] (7) -- (26);
\draw[black] (5) -- (0);
\draw[dotted] (5) -- (6);
\draw[dotted] (5) -- (21);
\draw[dotted] (5) -- (3);
\draw[dotted] (19) -- (10);
\draw[dotted] (19) -- (20);
\draw[dotted] (8) -- (24);
\draw[dotted] (23) -- (9);
\draw[black] (23) -- (0);
\draw[dotted] (9) -- (24);
\draw[black] (13) -- (1);
\draw[dotted] (13) -- (14);
\draw[black] (13) -- (22);
\draw[dotted] (13) -- (9);
\draw[black] (13) -- (20);
\draw[black] (13) -- (15);
\draw[dotted] (9) -- (0);
\draw[dotted] (9) -- (6);
\draw[black] (13) -- (2);
\draw[black] (13) -- (8);
\draw[black] (13) -- (26);
\draw[black] (20) -- (11);
\draw[dotted] (0) -- (14);
\draw[black] (0) -- (3);
\draw[black] (20) -- (5);
\draw[black] (20) -- (6);
\draw[black] (6) -- (7);
\draw[black] (7) -- (10);
\draw[black] (8) -- (23);
\draw[black] (22) -- (14);
\draw[black] (22) -- (8);
\draw[dotted] (8) -- (9);
\draw[black] (2) -- (1);
\draw[black] (1) -- (5);
\end{tikzpicture}
\begin{tikzpicture}[every node/.style={circle,draw},scale=0.24]
\node (14) at (0.3438825,8.89985545) {};
\node (26) at (0.819657165,5.13372485) {};
\node (6) at (15.30806175,6.045278) {};
\node (10) at (1.741297535,12.42932435) {};
\node (3) at (15.30806175,9.84121175) {};
\node (13) at (8.3949654,15.56024225) {};
\node (15) at (14.6038962,11.6200542) {};
\node (8) at (6.48562395,0.44637712) {};
\node (4) at (4.666221,1.037555225) {};
\node (7) at (10.274151,0.68473178) {};
\node (25) at (16.40848575,11.54792705) {};
\node (11) at (4.666221,14.8490227) {};
\node (12) at (13.47931225,13.1677018) {};
\node (27) at (0.819657165,10.7527649) {};
\node (0) at (14.6038962,4.26643555) {};
\node (17) at (3.050855,2.0625896) {};
\node (18) at (1.741297535,3.4571654) {};
\node (2) at (10.274151,15.2017227) {};
\node (16) at (15.49569815,13.20279545) {};
\node (23) at (12.0052026,1.49929243) {};
\node (1) at (12.0052026,14.38716205) {};
\node (9) at (13.47931225,2.71878795) {};
\node (24) at (8.3949654,0.3262475) {};
\node (5) at (15.54789775,7.94333305) {};
\node (21) at (17.12940455,7.94333305) {};
\node (19) at (3.050855,13.82390015) {};
\node (20) at (6.48562395,15.4401479) {};
\node (22) at (0.3438825,6.9866343) {};
\draw[dotted] (14) -- (26);
\draw[black] (14) -- (6);
\draw[black] (14) -- (10);
\draw[black] (3) -- (13);
\draw[black] (3) -- (15);
\draw[dotted] (3) -- (8);
\draw[dotted] (3) -- (4);
\draw[black] (3) -- (7);
\draw[black] (15) -- (25);
\draw[dotted] (4) -- (8);
\draw[dotted] (4) -- (26);
\draw[black] (15) -- (6);
\draw[dotted] (11) -- (10);
\draw[black] (12) -- (13);
\draw[black] (12) -- (15);
\draw[black] (14) -- (27);
\draw[dotted] (6) -- (0);
\draw[black] (27) -- (10);
\draw[black] (17) -- (18);
\draw[black] (18) -- (2);
\draw[black] (15) -- (16);
\draw[dotted] (4) -- (23);
\draw[black] (17) -- (1);
\draw[black] (17) -- (7);
\draw[black] (17) -- (8);
\draw[black] (17) -- (14);
\draw[dotted] (17) -- (15);
\draw[dotted] (7) -- (8);
\draw[dotted] (7) -- (23);
\draw[black] (7) -- (9);
\draw[black] (7) -- (24);
\draw[dotted] (7) -- (26);
\draw[black] (5) -- (0);
\draw[black] (5) -- (6);
\draw[black] (5) -- (21);
\draw[black] (5) -- (3);
\draw[black] (19) -- (10);
\draw[black] (19) -- (20);
\draw[black] (8) -- (24);
\draw[dotted] (23) -- (9);
\draw[dotted] (23) -- (0);
\draw[black] (9) -- (24);
\draw[black] (13) -- (1);
\draw[black] (13) -- (14);
\draw[dotted] (13) -- (22);
\draw[black] (13) -- (9);
\draw[black] (13) -- (20);
\draw[black] (13) -- (15);
\draw[black] (9) -- (0);
\draw[black] (9) -- (6);
\draw[black] (13) -- (2);
\draw[black] (13) -- (8);
\draw[dotted] (13) -- (26);
\draw[dotted] (20) -- (11);
\draw[black] (0) -- (14);
\draw[black] (0) -- (3);
\draw[black] (20) -- (5);
\draw[black] (20) -- (6);
\draw[black] (6) -- (7);
\draw[black] (7) -- (10);
\draw[dotted] (8) -- (23);
\draw[dotted] (22) -- (14);
\draw[dotted] (22) -- (8);
\draw[dotted] (8) -- (9);
\draw[black] (2) -- (1);
\draw[black] (1) -- (5);
\end{tikzpicture}
\end{center}
\caption{Links monitored by each of the four controllers as suggested by
algorithm \ref{alg:pathpart}. Critical links are more likely to be included in
multiple controllers, whereas less important links are appeared in only one
controller.}\label{fig:ebone2}
\end{figure}

We use $q=4$ controllers on a network of 28 nodes and 66 links. The topology is
illustrated in \figurename~\ref{fig:ebone}. One configuration of the four
controllers computed by algorithm \ref{alg:pathpart} is depicted in
\figurename~\ref{fig:ebone2}, with each controller monitoring 45--47 links.
From the figure, we found quite significant overlap on the nodes and links
monitored by each controller. Some links appeared in all controllers, as they
are critical links for the connection of the network. Some other links are less
important and appeared in one controller only. The large number of overlap is
unavoidable when devolved controllers are used. In fact,
\begin{theorem}
When devolved controllers are used, there is either a controller that
covers all nodes, or any single node is covered by more than one controller.
\end{theorem}
Since for any node $v$, if it is covered by only one controller, then for any
flow $(v,u)$ to be routable, node $u$ must also covered by that controller.
Therefore that controller must cover all nodes on the network. $\Box$

In other words, there must be significant overlap if we want to reduce the
scope of the network that each controller monitors. In fact, we can reduce the
number of links monitored by each controller if we use Algorithm
\ref{alg:partpath}. Applying to the same topology in
\figurename~\ref{fig:ebone}, each controller monitors only 29--31 links, which
is significantly less. This better result, however, comes with the price that
the route found by Algorithm \ref{alg:partpath} is longer. The average hop
count of a path (the mean number of hops over all $kn(n-1)$ paths computed) is
3.5 in Algorithm \ref{alg:partpath} whereas that in Algorithm
\ref{alg:pathpart} is 2.6. The lengthened route may not be favorable in data
center networks, however.

\begin{figure}
\centering
\includegraphics[scale=0.8]{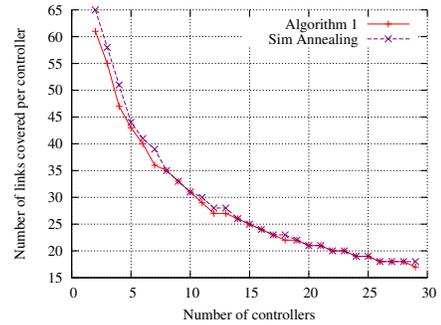}
\caption{Number of links covered per controller vs number of controllers in topology of \figurename~\ref{fig:ebone}, comparing Algorithm \ref{alg:pathpart} and simulated annealing}\label{fig:anneal}
\end{figure}

To compare the result, we computed a configuration with the same set of
multipaths using the time-consuming \emph{simulated annealing}
process\footnote{The simulated annealing process is to replace the loop in
lines \ref{alg:partition1}--\ref{alg:partition2} only in Algorithm
\ref{alg:pathpart}. The multipath $\mathcal{M}$ in the comparison are exactly
the same for a fair contrast.  It may seem counter-intuitive that the
well-established simulated annealing technique does not produce better
solution. Partly this can be attributed to the choice of parameters such as the
cooling function used. More importantly, however, is because our algorithm work
`smarter' than simulated annealing as our cost function guides toward
optimality whereas the latter is simply a brute-force search.}.  The result,
presented in \figurename~\ref{fig:anneal}, turns out is no better than that
obtained by the heuristic algorithms despite the longer time it took.  Indeed,
the heuristic algorithms often gives a slightly better solution than simulated
annealing.

We also applied the algorithm on several different topologies of different
number of nodes and links from Rocketfuel. Due to space limitation, we do not
show their topologies here but \tablename~\ref{tab:topos} shows the maximum
number of links covered by a controller resulted from the heuristic algorithm
compared to that from simulated annealing. It confirms that the controllers
covers around 60-80\% of links on the network when $q=4$ and the result
provided by Algorithm \ref{alg:pathpart} is at least as good as that obtained
by simulated annealing.

\begin{table}
   \centering
   \begin{tabular}{l|ccc@{ }cc@{ }c}
      Topology & \# nodes & \# links & \multicolumn{2}{c}{Algo. 1} & \multicolumn{2}{c}{Sim. Annealing} \\
      \hline
      1 (Fig.\ref{fig:ebone}) & 28 & 66 & 47 & (0.1s) & 51 & (69.9s)\\ % ebone
      2 & 108 & 141 & 114 & (24.4s) & 140 & (1387.4s) \\ % att
      3 & 53 & 456 & 204 & (1.1s) & 226 & (301.6s) \\ % level 3
      4 & 44 & 106 & 77 & (0.5s) & 92 & (186.6s) \\ % sprintlink
      5 & 51 & 129 & 97 & (0.9s) & 112 & (239.9s) \\% tiscali
      6 (Fig.\ref{fig:fattree}) & 45 & 108 & 83 & (0.03s) & 83 & (46.3s) % fat tree
   \end{tabular}
   \caption{Comparing the size of controllers in different topologies, with time taken by each solver shown in brackets.}
   \label{tab:topos}
\end{table}

\subsection{Number of controllers and the effect on the size of coverage}

\begin{figure}
  \centering
  \includegraphics[scale=0.8]{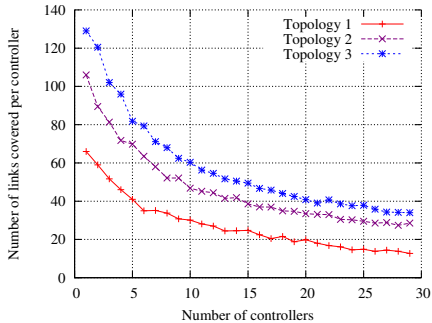}
  \caption{Number of links covered per controller vs number of controllers in three different topologies from Rocketfuel}\label{fig:edge}
\end{figure}

In order to reduce the number of links covered by any controller, an intuitive
way is to use more controllers. We applied Algorithm \ref{alg:pathpart} with
various $q$ to three different irregular topologies. \figurename~\ref{fig:edge}
plots the result.

Obviously, $q=1$ shows the total number of links in the network.  In
\figurename~\ref{fig:edge}, the curves show a general decreasing trend. In
fact, the curves are decreasing geometrically. This suggests that although we
can reduce the size of a controller, there is an overhead: As $q$ increases,
the average number of controllers monitoring a link also increases. This means
the monitoring traffic, although small, also increases with $q$. This is the
trade-off that we have to consider when devolved controllers are used in place
of a single centralized controller.

\subsection{The effect on the number of paths}

\begin{figure}
\centering
\includegraphics[scale=0.8]{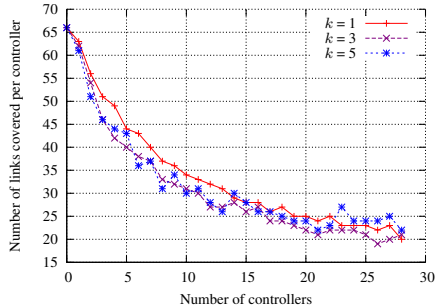}
\caption{Number of links covered per controller vs number of controllers in topology of \figurename~\ref{fig:ebone} with different value of $k$}\label{fig:k}
\end{figure}

While the parameter $q$ affects the number of links covered by each controller,
the parameter $k$, i.e. the number of paths to find between a
source-destination pair, does not have a significant effect on it. This is
shown by \figurename~\ref{fig:k}. The figure plots the number of links covered
by a controller against the number of controllers using the topology of
\figurename~\ref{fig:ebone}, but with the parameter $k$ varied. We examined
with $k$ in ranges of 1 to 10 and also some larger values. The varied value of
$k$ does not produce a significantly different result between each other. This
is explained by the fact that each link on the network is reused for fairly
large number of times in different flows $(s,t)$. When we pick a multipath from
a controller, it is likely that every links on this multipath are also used by
another multipath from the same controller. In this sense, if we increase the
multiplicity $k$, the additional paths also likely using the links already
covered by the controller.

\subsection{Using partition-path approach on regular networks}\label{sub:regular}

As mentioned in section \ref{sub:size}, Algorithm \ref{alg:partpath} produces a
better result because it has a path-finding algorithm that fits the path into
the controller, in the expense of resulting in a longer route.  This weakness
of partition-path algorithm can be removed on regular topology networks such as
fat tree, which according to \cite{alv08}, has been suggested to use in data
center networks. It is because in a regular topology, we know a priori the
length of an optimum route and it also provides enough number of distinct paths
of the \emph{same length} between a source-destination pair.

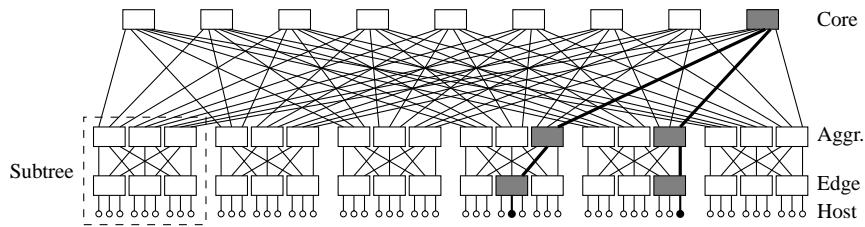
\begin{figure*}
\centering\input{fattree.tex}
\caption{A fat tree topology built with switches of 6 ports}\label{fig:fattree}
\end{figure*}

To illustrate the idea, we show a 3-layer fat tree network built with switches
of 6 ports in \figurename~\ref{fig:fattree}. It is trivial to see that, given a
pair of hosts in a different subtree, there are $(6/2)^2=9$ distinct paths
(each pass through a distinct core switch) between them that passes through 5
switches. In such a network, no path that traverses more than 5 switches is
optimal. With such knowledge, we can modify the path-finding algorithm used in
Algorithm \ref{alg:partpath} (line \ref{alg:dijkstra}) to enforce a solution of
fixed-length path. Note that such modification only works on regular topologies
like fat tree or Clos network. In fact, the modified path-finding algorithm can
be used in place of that in line \ref{alg:pathfind1} of Algorithm
\ref{alg:pathpart} as well.

We also modified line \ref{alg:partlink} in Algorithm \ref{alg:partpath}
slightly so that only the links connecting core and aggregation switches are
partitioned into $\mathcal{E}_i$. This is a reasonable modification considering
that in a fat tree network (see \figurename~\ref{fig:fattree}), we fixed the
whole path between two nodes when we fixed the links that it uses connecting
the core and aggregation switches.

Applying the path-partition and partition-path algorithms to the network in
\figurename~\ref{fig:fattree} with $q=4$, we find the coverage per
controller to be 83 and 56 links respectively over a total of 108 links.
Both Algorithm \ref{alg:pathpart} and Algorithm \ref{alg:partpath} yield an
average hope count of 3.8, due to the modified path-finding algorithm.

\section{Discussions}\label{sec:discussion}

\subsection{The communication between a server and controllers}

According to the algorithms aforementioned, the multipath for each $(s,t)$ is
installed in only one controller. Therefore, only one among the $q$ controllers
can reply to a route request for $(s,t)$. When node $s$ is initiating a flow to
$t$, it has to deduce which controller can answer its route request. There are
two ways to solve this problem. Firstly, node $s$ can send its request to all
$q$ controllers and let the one owns the data to reply. Trivially this solution
incurs additional network traffic. The second solution is to have a mapping
at node $s$: For each destination $t$, there is a table in $s$ tells which
controller contains the route for $(s,t)$. This is a viable solution because
the total number of nodes $|V|$ (and the number of destinations $t$ for any
node $s$) is usually limited. Moreover, when we configured the controllers,
storing the mapping information of $\{(s,t), \forall t\in V\}$ to $s$ is just
one step further with the existing information.

\subsection{Path-partition vs partition-path}

Section \ref{sub:size} mention that path-partition algorithm is inferior to
partition-path algorithm in terms of the size of controllers
produced. However, only path-partition algorithm can guarantee
shortest-path routes because the path-finding algorithm is not interfered by
the configuration of controllers.

When regular topologies are used, such as fat tree networks, we can have a
modified path-finding algorithm to ensure shortest-path routes are found. This
makes the partition-path algorithm favorable. Therefore, it is interesting to
see that path-partition algorithm is suitable for use with irregular networks
while partition-path algorithm is good for regular networks.

\subsection{Precompute routes}

Using pre-computed routes in this paper is intentional. Assume each
controller covers only a part of the network and a flow's route is computed
dynamically when the request is arrived. It is hard to guarantee that, among
the $q$ controllers, there must be one can fulfill any route request. The role
of pre-computed multipaths is therefore a verifier to guarantee a controller is
responsible for any possible flow.

\subsection{Link failures}

While we do not address the actual operation of devolved controllers in a
network, it is expected that whenever there is a link failure, i.e. a topology
change, something have to be done in the controllers to reflect this change.
This could be disabling certain paths (among the $k$ multiple paths of the same
source-destination pair), or reconfiguration of the network. This overhead
could be large and intensive. In order to provide a prompt reaction, therefore
it is essential to keeping the number of links managed by a controller small.
This justifies our objective in the optimization.

\subsection{Redundancy}\label{sec:redundancy}

\begin{figure}
\centering
\includegraphics[scale=0.8]{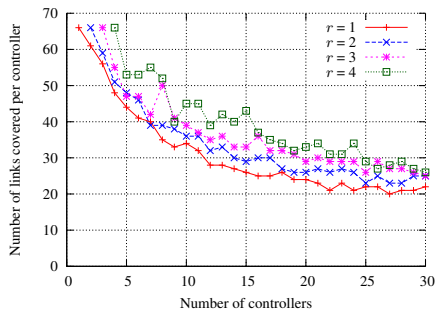}
\caption{Number of links covered per controller vs number of controllers in
topology of \figurename~\ref{fig:ebone} with different value of $r$}\label{fig:redun}
\end{figure}

While it is possible for more than one controller that can respond to a route
request, the algorithms in section \ref{sec:algo} does not guarantee this.  One
way to ensure redundancy is to modify line \ref{alg:partition2} of algorithm
\ref{alg:pathpart} or line \ref{alg:partition} of algorithm \ref{alg:partpath},
so that a path is added to $r>1$ controllers.  Usually $r=2$ is sufficient for
resilience.  In \figurename~\ref{fig:redun}, we plot the number of links
covered by each controller in different values of $r$ in topology of
\figurename~\ref{fig:ebone}, using the modified path-partition algorithm.
Trivially, as the degree of redundancy $r$ increases, the number of links
covered by a controller increases. The increment, however, is moderate due to
the overlap of link coverage between controllers. In other words, we can have
redundancy at just a small price. More details on the redundancy design, its
overhead, and an example showing its mechanism would be in a future
paper.

% Figure: Topology 1, $r$=1,2,3,4,5
% Alternative approach for redundancy

\section{Conclusion}\label{sec:conclu}

The focus of this paper is to see the possibility of using multiple small
independent controllers instead of a single centralized omniscient controller
to manage resources. We use flow routing as an example to see how we can use
multiple controllers to assign routes to flows base on dynamic network status.
The main reason to avoid a single controller is because of scalability concern.
Therefore, we forbid our controllers to have the complete network topology
information in run time, and introduced the concept of \emph{devolved
controllers}. Furthermore, we propose algorithms that aims to limit the network
topology information stored in the controllers.

Our result shows that, devolved controllers are possible. We proposed two
heuristic algorithms to limit the size of each controller. Although they do not
seek for a globally optimal solution, their results are as good as simulated
annealing solvers but much faster. The heuristic algorithms, path-partition and
partition-path algorithms, are found to be suitable for irregular and regular
networks respectively. Such difference is due to the fact that, in regular
networks, we can easily estimate the length of route a priori.

In computer networks such as data center or compute clouds, controllers are
often used, such as for security policy control, resource allocation, billing,
and so on. This paper is a precursor to a new design direction on the use of
controllers, such that they can scale out.

\bibliographystyle{IEEEtran}
\bibliography{control}

\label{docend}
\end{document}

%% file: fattree.tex
\begin{tikzpicture}[scale=0.26]
\draw (0,0) rectangle +(1.618,1);	% Pod 0 switch 0 edge
\draw (0,2.5) rectangle (1.618,3.5);		% Pod 0 switch 0 agg
\draw (1.818,0) rectangle +(1.618,1);	% Pod 0 switch 1 edge
\draw (1.818,2.5) rectangle (3.436,3.5);		% Pod 0 switch 1 agg
\draw (3.636,0) rectangle +(1.618,1);	% Pod 0 switch 2 edge
\draw (3.636,2.5) rectangle (5.254,3.5);		% Pod 0 switch 2 agg
\draw (6.254,0) rectangle +(1.618,1);	% Pod 1 switch 0 edge
\draw (6.254,2.5) rectangle (7.872,3.5);		% Pod 1 switch 0 agg
\draw (8.072,0) rectangle +(1.618,1);	% Pod 1 switch 1 edge
\draw (8.072,2.5) rectangle (9.69,3.5);		% Pod 1 switch 1 agg
\draw (9.89,0) rectangle +(1.618,1);	% Pod 1 switch 2 edge
\draw (9.89,2.5) rectangle (11.508,3.5);		% Pod 1 switch 2 agg
\draw (12.508,0) rectangle +(1.618,1);	% Pod 2 switch 0 edge
\draw (12.508,2.5) rectangle (14.126,3.5);		% Pod 2 switch 0 agg
\draw (14.326,0) rectangle +(1.618,1);	% Pod 2 switch 1 edge
\draw (14.326,2.5) rectangle (15.944,3.5);		% Pod 2 switch 1 agg
\draw (16.144,0) rectangle +(1.618,1);	% Pod 2 switch 2 edge
\draw (16.144,2.5) rectangle (17.762,3.5);		% Pod 2 switch 2 agg
\draw (18.762,0) rectangle +(1.618,1);	% Pod 3 switch 0 edge
\draw (18.762,2.5) rectangle (20.38,3.5);		% Pod 3 switch 0 agg
\draw[fill=gray] (20.58,0) rectangle +(1.618,1);	% Pod 3 switch 1 edge
\draw (20.58,2.5) rectangle (22.198,3.5);		% Pod 3 switch 1 agg
\draw (22.398,0) rectangle +(1.618,1);	% Pod 3 switch 2 edge
\draw[fill=gray] (22.398,2.5) rectangle (24.016,3.5);		% Pod 3 switch 2 agg
\draw (25.016,0) rectangle +(1.618,1);	% Pod 4 switch 0 edge
\draw (25.016,2.5) rectangle (26.634,3.5);		% Pod 4 switch 0 agg
\draw (26.834,0) rectangle +(1.618,1);	% Pod 4 switch 1 edge
\draw (26.834,2.5) rectangle (28.452,3.5);		% Pod 4 switch 1 agg
\draw[fill=gray] (28.652,0) rectangle +(1.618,1);	% Pod 4 switch 2 edge
\draw[fill=gray] (28.652,2.5) rectangle (30.27,3.5);		% Pod 4 switch 2 agg
\draw (31.27,0) rectangle +(1.618,1);	% Pod 5 switch 0 edge
\draw (31.27,2.5) rectangle (32.888,3.5);		% Pod 5 switch 0 agg
\draw (33.088,0) rectangle +(1.618,1);	% Pod 5 switch 1 edge
\draw (33.088,2.5) rectangle (34.706,3.5);		% Pod 5 switch 1 agg
\draw (34.906,0) rectangle +(1.618,1);	% Pod 5 switch 2 edge
\draw (34.906,2.5) rectangle (36.524,3.5);		% Pod 5 switch 2 agg

\tikzstyle{every node}=[font=\footnotesize]
\node[text width=0cm] at (37,-0.8) {Host};
\node[text width=0cm] at (37,0.5) {Edge};
\node[text width=0cm] at (37,3.0) {Aggr.};
\node[text width=0cm] at (37,9.0) {Core};

\node[anchor=east] at (-0.5,1.25) {Subtree};
\draw[dashed] (-0.5,-1.5) rectangle (5.75,4);

\draw (0.269666666666667,1) -- (0.269666666666667,2.5);	% Pod 0 edge 0 output 0
\draw (0.809,1) -- (2.08766666666667,2.5);	% Pod 0 edge 0 output 1
\draw (1.34833333333333,1) -- (3.90566666666667,2.5);	% Pod 0 edge 0 output 2
\draw (2.08766666666667,1) -- (0.809,2.5);	% Pod 0 edge 1 output 0
\draw (2.627,1) -- (2.627,2.5);	% Pod 0 edge 1 output 1
\draw (3.16633333333333,1) -- (4.445,2.5);	% Pod 0 edge 1 output 2
\draw (3.90566666666667,1) -- (1.34833333333333,2.5);	% Pod 0 edge 2 output 0
\draw (4.445,1) -- (3.16633333333333,2.5);	% Pod 0 edge 2 output 1
\draw (4.98433333333333,1) -- (4.98433333333333,2.5);	% Pod 0 edge 2 output 2
\draw (6.52366666666667,1) -- (6.52366666666667,2.5);	% Pod 1 edge 0 output 0
\draw (7.063,1) -- (8.34166666666667,2.5);	% Pod 1 edge 0 output 1
\draw (7.60233333333333,1) -- (10.1596666666667,2.5);	% Pod 1 edge 0 output 2
\draw (8.34166666666667,1) -- (7.063,2.5);	% Pod 1 edge 1 output 0
\draw (8.881,1) -- (8.881,2.5);	% Pod 1 edge 1 output 1
\draw (9.42033333333333,1) -- (10.699,2.5);	% Pod 1 edge 1 output 2
\draw (10.1596666666667,1) -- (7.60233333333333,2.5);	% Pod 1 edge 2 output 0
\draw (10.699,1) -- (9.42033333333333,2.5);	% Pod 1 edge 2 output 1
\draw (11.2383333333333,1) -- (11.2383333333333,2.5);	% Pod 1 edge 2 output 2
\draw (12.7776666666667,1) -- (12.7776666666667,2.5);	% Pod 2 edge 0 output 0
\draw (13.317,1) -- (14.5956666666667,2.5);	% Pod 2 edge 0 output 1
\draw (13.8563333333333,1) -- (16.4136666666667,2.5);	% Pod 2 edge 0 output 2
\draw (14.5956666666667,1) -- (13.317,2.5);	% Pod 2 edge 1 output 0
\draw (15.135,1) -- (15.135,2.5);	% Pod 2 edge 1 output 1
\draw (15.6743333333333,1) -- (16.953,2.5);	% Pod 2 edge 1 output 2
\draw (16.4136666666667,1) -- (13.8563333333333,2.5);	% Pod 2 edge 2 output 0
\draw (16.953,1) -- (15.6743333333333,2.5);	% Pod 2 edge 2 output 1
\draw (17.4923333333333,1) -- (17.4923333333333,2.5);	% Pod 2 edge 2 output 2
\draw (19.0316666666667,1) -- (19.0316666666667,2.5);	% Pod 3 edge 0 output 0
\draw (19.571,1) -- (20.8496666666667,2.5);	% Pod 3 edge 0 output 1
\draw (20.1103333333333,1) -- (22.6676666666667,2.5);	% Pod 3 edge 0 output 2
\draw (20.8496666666667,1) -- (19.571,2.5);	% Pod 3 edge 1 output 0
\draw (21.389,1) -- (21.389,2.5);	% Pod 3 edge 1 output 1
\draw[very thick] (21.9283333333333,1) -- (23.207,2.5);	% Pod 3 edge 1 output 2
\draw (22.6676666666667,1) -- (20.1103333333333,2.5);	% Pod 3 edge 2 output 0
\draw (23.207,1) -- (21.9283333333333,2.5);	% Pod 3 edge 2 output 1
\draw (23.7463333333333,1) -- (23.7463333333333,2.5);	% Pod 3 edge 2 output 2
\draw (25.2856666666667,1) -- (25.2856666666667,2.5);	% Pod 4 edge 0 output 0
\draw (25.825,1) -- (27.1036666666667,2.5);	% Pod 4 edge 0 output 1
\draw (26.3643333333333,1) -- (28.9216666666667,2.5);	% Pod 4 edge 0 output 2
\draw (27.1036666666667,1) -- (25.825,2.5);	% Pod 4 edge 1 output 0
\draw (27.643,1) -- (27.643,2.5);	% Pod 4 edge 1 output 1
\draw (28.1823333333333,1) -- (29.461,2.5);	% Pod 4 edge 1 output 2
\draw (28.9216666666667,1) -- (26.3643333333333,2.5);	% Pod 4 edge 2 output 0
\draw (29.461,1) -- (28.1823333333333,2.5);	% Pod 4 edge 2 output 1
\draw[very thick] (30.0003333333333,1) -- (30.0003333333333,2.5);	% Pod 4 edge 2 output 2
\draw (31.5396666666667,1) -- (31.5396666666667,2.5);	% Pod 5 edge 0 output 0
\draw (32.079,1) -- (33.3576666666667,2.5);	% Pod 5 edge 0 output 1
\draw (32.6183333333333,1) -- (35.1756666666667,2.5);	% Pod 5 edge 0 output 2
\draw (33.3576666666667,1) -- (32.079,2.5);	% Pod 5 edge 1 output 0
\draw (33.897,1) -- (33.897,2.5);	% Pod 5 edge 1 output 1
\draw (34.4363333333333,1) -- (35.715,2.5);	% Pod 5 edge 1 output 2
\draw (35.1756666666667,1) -- (32.6183333333333,2.5);	% Pod 5 edge 2 output 0
\draw (35.715,1) -- (34.4363333333333,2.5);	% Pod 5 edge 2 output 1
\draw (36.2543333333333,1) -- (36.2543333333333,2.5);	% Pod 5 edge 2 output 2

\draw (1.5,8.5) rectangle +(1.618,1);		% Core 0
\draw (5.48825,8.5) rectangle +(1.618,1);		% Core 1
\draw (9.4765,8.5) rectangle +(1.618,1);		% Core 2
\draw (13.46475,8.5) rectangle +(1.618,1);		% Core 3
\draw (17.453,8.5) rectangle +(1.618,1);		% Core 4
\draw (21.44125,8.5) rectangle +(1.618,1);		% Core 5
\draw (25.4295,8.5) rectangle +(1.618,1);		% Core 6
\draw (29.41775,8.5) rectangle +(1.618,1);		% Core 7
\draw[fill=gray] (33.406,8.5) rectangle +(1.618,1);		% Core 8

\draw (0.269666666666667,3.5) -- (1.63483333333333,8.5);	% Pod 0 switch 0 to core 0
\draw (0.809,3.5) -- (5.62308333333333,8.5);	% Pod 0 switch 0 to core 1
\draw (1.34833333333333,3.5) -- (9.61133333333333,8.5);	% Pod 0 switch 0 to core 2
\draw (2.08766666666667,3.5) -- (13.5995833333333,8.5);	% Pod 0 switch 1 to core 3
\draw (2.627,3.5) -- (17.5878333333333,8.5);	% Pod 0 switch 1 to core 4
\draw (3.16633333333333,3.5) -- (21.5760833333333,8.5);	% Pod 0 switch 1 to core 5
\draw (3.90566666666667,3.5) -- (25.5643333333333,8.5);	% Pod 0 switch 2 to core 6
\draw (4.445,3.5) -- (29.5525833333333,8.5);	% Pod 0 switch 2 to core 7
\draw (4.98433333333333,3.5) -- (33.5408333333333,8.5);	% Pod 0 switch 2 to core 8
\draw (6.52366666666667,3.5) -- (1.90450000000001,8.5);	% Pod 1 switch 0 to core 0
\draw (7.063,3.5) -- (5.89275,8.5);	% Pod 1 switch 0 to core 1
\draw (7.60233333333333,3.5) -- (9.881,8.5);	% Pod 1 switch 0 to core 2
\draw (8.34166666666667,3.5) -- (13.86925,8.5);	% Pod 1 switch 1 to core 3
\draw (8.881,3.5) -- (17.8575,8.5);	% Pod 1 switch 1 to core 4
\draw (9.42033333333333,3.5) -- (21.84575,8.5);	% Pod 1 switch 1 to core 5
\draw (10.1596666666667,3.5) -- (25.834,8.5);	% Pod 1 switch 2 to core 6
\draw (10.699,3.5) -- (29.82225,8.5);	% Pod 1 switch 2 to core 7
\draw (11.2383333333333,3.5) -- (33.8105,8.5);	% Pod 1 switch 2 to core 8
\draw (12.7776666666667,3.5) -- (2.17416666666668,8.5);	% Pod 2 switch 0 to core 0
\draw (13.317,3.5) -- (6.16241666666668,8.5);	% Pod 2 switch 0 to core 1
\draw (13.8563333333333,3.5) -- (10.1506666666667,8.5);	% Pod 2 switch 0 to core 2
\draw (14.5956666666667,3.5) -- (14.1389166666667,8.5);	% Pod 2 switch 1 to core 3
\draw (15.135,3.5) -- (18.1271666666667,8.5);	% Pod 2 switch 1 to core 4
\draw (15.6743333333333,3.5) -- (22.1154166666667,8.5);	% Pod 2 switch 1 to core 5
\draw (16.4136666666667,3.5) -- (26.1036666666667,8.5);	% Pod 2 switch 2 to core 6
\draw (16.953,3.5) -- (30.0919166666667,8.5);	% Pod 2 switch 2 to core 7
\draw (17.4923333333333,3.5) -- (34.0801666666667,8.5);	% Pod 2 switch 2 to core 8
\draw (19.0316666666667,3.5) -- (2.44383333333335,8.5);	% Pod 3 switch 0 to core 0
\draw (19.571,3.5) -- (6.43208333333335,8.5);	% Pod 3 switch 0 to core 1
\draw (20.1103333333333,3.5) -- (10.4203333333333,8.5);	% Pod 3 switch 0 to core 2
\draw (20.8496666666667,3.5) -- (14.4085833333333,8.5);	% Pod 3 switch 1 to core 3
\draw (21.389,3.5) -- (18.3968333333333,8.5);	% Pod 3 switch 1 to core 4
\draw (21.9283333333333,3.5) -- (22.3850833333333,8.5);	% Pod 3 switch 1 to core 5
\draw (22.6676666666666,3.5) -- (26.3733333333333,8.5);	% Pod 3 switch 2 to core 6
\draw (23.207,3.5) -- (30.3615833333333,8.5);	% Pod 3 switch 2 to core 7
\draw[very thick] (23.7463333333333,3.5) -- (34.3498333333334,8.5);	% Pod 3 switch 2 to core 8
\draw (25.2856666666666,3.5) -- (2.71350000000002,8.5);	% Pod 4 switch 0 to core 0
\draw (25.825,3.5) -- (6.70175000000002,8.5);	% Pod 4 switch 0 to core 1
\draw (26.3643333333333,3.5) -- (10.69,8.5);	% Pod 4 switch 0 to core 2
\draw (27.1036666666666,3.5) -- (14.67825,8.5);	% Pod 4 switch 1 to core 3
\draw (27.643,3.5) -- (18.6665,8.5);	% Pod 4 switch 1 to core 4
\draw (28.1823333333333,3.5) -- (22.65475,8.5);	% Pod 4 switch 1 to core 5
\draw (28.9216666666666,3.5) -- (26.643,8.5);	% Pod 4 switch 2 to core 6
\draw (29.461,3.5) -- (30.63125,8.5);	% Pod 4 switch 2 to core 7
\draw[very thick] (30.0003333333333,3.5) -- (34.6195,8.5);	% Pod 4 switch 2 to core 8
\draw (31.5396666666666,3.5) -- (2.9831666666667,8.5);	% Pod 5 switch 0 to core 0
\draw (32.079,3.5) -- (6.9714166666667,8.5);	% Pod 5 switch 0 to core 1
\draw (32.6183333333333,3.5) -- (10.9596666666667,8.5);	% Pod 5 switch 0 to core 2
\draw (33.3576666666666,3.5) -- (14.9479166666667,8.5);	% Pod 5 switch 1 to core 3
\draw (33.897,3.5) -- (18.9361666666667,8.5);	% Pod 5 switch 1 to core 4
\draw (34.4363333333333,3.5) -- (22.9244166666667,8.5);	% Pod 5 switch 1 to core 5
\draw (35.1756666666666,3.5) -- (26.9126666666667,8.5);	% Pod 5 switch 2 to core 6
\draw (35.715,3.5) -- (30.9009166666667,8.5);	% Pod 5 switch 2 to core 7
\draw (36.2543333333333,3.5) -- (34.8891666666667,8.5);	% Pod 5 switch 2 to core 8

\draw (0.269666666666667,0) -- (0.269666666666667,-0.8)		% Pod 0 switch 0 host 0
(0.269666666666667,-0.95) circle (0.15);
\draw (0.809,0) -- (0.809,-0.8)		% Pod 0 switch 0 host 1
(0.809,-0.95) circle (0.15);
\draw (1.34833333333333,0) -- (1.34833333333333,-0.8)		% Pod 0 switch 0 host 2
(1.34833333333333,-0.95) circle (0.15);
\draw (2.08766666666667,0) -- (2.08766666666667,-0.8)		% Pod 0 switch 1 host 0
(2.08766666666667,-0.95) circle (0.15);
\draw (2.627,0) -- (2.627,-0.8)		% Pod 0 switch 1 host 1
(2.627,-0.95) circle (0.15);
\draw (3.16633333333333,0) -- (3.16633333333333,-0.8)		% Pod 0 switch 1 host 2
(3.16633333333333,-0.95) circle (0.15);
\draw (3.90566666666667,0) -- (3.90566666666667,-0.8)		% Pod 0 switch 2 host 0
(3.90566666666667,-0.95) circle (0.15);
\draw (4.445,0) -- (4.445,-0.8)		% Pod 0 switch 2 host 1
(4.445,-0.95) circle (0.15);
\draw (4.98433333333333,0) -- (4.98433333333333,-0.8)		% Pod 0 switch 2 host 2
(4.98433333333333,-0.95) circle (0.15);
\draw (6.52366666666667,0) -- (6.52366666666667,-0.8)		% Pod 1 switch 0 host 0
(6.52366666666667,-0.95) circle (0.15);
\draw (7.063,0) -- (7.063,-0.8)		% Pod 1 switch 0 host 1
(7.063,-0.95) circle (0.15);
\draw (7.60233333333333,0) -- (7.60233333333333,-0.8)		% Pod 1 switch 0 host 2
(7.60233333333333,-0.95) circle (0.15);
\draw (8.34166666666667,0) -- (8.34166666666667,-0.8)		% Pod 1 switch 1 host 0
(8.34166666666667,-0.95) circle (0.15);
\draw (8.881,0) -- (8.881,-0.8)		% Pod 1 switch 1 host 1
(8.881,-0.95) circle (0.15);
\draw (9.42033333333333,0) -- (9.42033333333333,-0.8)		% Pod 1 switch 1 host 2
(9.42033333333333,-0.95) circle (0.15);
\draw (10.1596666666667,0) -- (10.1596666666667,-0.8)		% Pod 1 switch 2 host 0
(10.1596666666667,-0.95) circle (0.15);
\draw (10.699,0) -- (10.699,-0.8)		% Pod 1 switch 2 host 1
(10.699,-0.95) circle (0.15);
\draw (11.2383333333333,0) -- (11.2383333333333,-0.8)		% Pod 1 switch 2 host 2
(11.2383333333333,-0.95) circle (0.15);
\draw (12.7776666666667,0) -- (12.7776666666667,-0.8)		% Pod 2 switch 0 host 0
(12.7776666666667,-0.95) circle (0.15);
\draw (13.317,0) -- (13.317,-0.8)		% Pod 2 switch 0 host 1
(13.317,-0.95) circle (0.15);
\draw (13.8563333333333,0) -- (13.8563333333333,-0.8)		% Pod 2 switch 0 host 2
(13.8563333333333,-0.95) circle (0.15);
\draw (14.5956666666667,0) -- (14.5956666666667,-0.8)		% Pod 2 switch 1 host 0
(14.5956666666667,-0.95) circle (0.15);
\draw (15.135,0) -- (15.135,-0.8)		% Pod 2 switch 1 host 1
(15.135,-0.95) circle (0.15);
\draw (15.6743333333333,0) -- (15.6743333333333,-0.8)		% Pod 2 switch 1 host 2
(15.6743333333333,-0.95) circle (0.15);
\draw (16.4136666666667,0) -- (16.4136666666667,-0.8)		% Pod 2 switch 2 host 0
(16.4136666666667,-0.95) circle (0.15);
\draw (16.953,0) -- (16.953,-0.8)		% Pod 2 switch 2 host 1
(16.953,-0.95) circle (0.15);
\draw (17.4923333333333,0) -- (17.4923333333333,-0.8)		% Pod 2 switch 2 host 2
(17.4923333333333,-0.95) circle (0.15);
\draw (19.0316666666667,0) -- (19.0316666666667,-0.8)		% Pod 3 switch 0 host 0
(19.0316666666667,-0.95) circle (0.15);
\draw (19.571,0) -- (19.571,-0.8)		% Pod 3 switch 0 host 1
(19.571,-0.95) circle (0.15);
\draw (20.1103333333333,0) -- (20.1103333333333,-0.8)		% Pod 3 switch 0 host 2
(20.1103333333333,-0.95) circle (0.15);
\draw (20.8496666666667,0) -- (20.8496666666667,-0.8)		% Pod 3 switch 1 host 0
(20.8496666666667,-0.95) circle (0.15);
\draw[fill,thick] (21.389,0) -- (21.389,-0.8)		% Pod 3 switch 1 host 1
(21.389,-0.95) circle (0.15);
\draw (21.9283333333333,0) -- (21.9283333333333,-0.8)		% Pod 3 switch 1 host 2
(21.9283333333333,-0.95) circle (0.15);
\draw (22.6676666666666,0) -- (22.6676666666666,-0.8)		% Pod 3 switch 2 host 0
(22.6676666666666,-0.95) circle (0.15);
\draw (23.207,0) -- (23.207,-0.8)		% Pod 3 switch 2 host 1
(23.207,-0.95) circle (0.15);
\draw (23.7463333333333,0) -- (23.7463333333333,-0.8)		% Pod 3 switch 2 host 2
(23.7463333333333,-0.95) circle (0.15);
\draw (25.2856666666666,0) -- (25.2856666666666,-0.8)		% Pod 4 switch 0 host 0
(25.2856666666666,-0.95) circle (0.15);
\draw (25.825,0) -- (25.825,-0.8)		% Pod 4 switch 0 host 1
(25.825,-0.95) circle (0.15);
\draw (26.3643333333333,0) -- (26.3643333333333,-0.8)		% Pod 4 switch 0 host 2
(26.3643333333333,-0.95) circle (0.15);
\draw (27.1036666666666,0) -- (27.1036666666666,-0.8)		% Pod 4 switch 1 host 0
(27.1036666666666,-0.95) circle (0.15);
\draw (27.643,0) -- (27.643,-0.8)		% Pod 4 switch 1 host 1
(27.643,-0.95) circle (0.15);
\draw (28.1823333333333,0) -- (28.1823333333333,-0.8)		% Pod 4 switch 1 host 2
(28.1823333333333,-0.95) circle (0.15);
\draw (28.9216666666666,0) -- (28.9216666666666,-0.8)		% Pod 4 switch 2 host 0
(28.9216666666666,-0.95) circle (0.15);
\draw (29.461,0) -- (29.461,-0.8)		% Pod 4 switch 2 host 1
(29.461,-0.95) circle (0.15);
\draw[fill,thick] (30.0003333333333,0) -- (30.0003333333333,-0.8)		% Pod 4 switch 2 host 2
(30.0003333333333,-0.95) circle (0.15);
\draw (31.5396666666666,0) -- (31.5396666666666,-0.8)		% Pod 5 switch 0 host 0
(31.5396666666666,-0.95) circle (0.15);
\draw (32.079,0) -- (32.079,-0.8)		% Pod 5 switch 0 host 1
(32.079,-0.95) circle (0.15);
\draw (32.6183333333333,0) -- (32.6183333333333,-0.8)		% Pod 5 switch 0 host 2
(32.6183333333333,-0.95) circle (0.15);
\draw (33.3576666666666,0) -- (33.3576666666666,-0.8)		% Pod 5 switch 1 host 0
(33.3576666666666,-0.95) circle (0.15);
\draw (33.897,0) -- (33.897,-0.8)		% Pod 5 switch 1 host 1
(33.897,-0.95) circle (0.15);
\draw (34.4363333333333,0) -- (34.4363333333333,-0.8)		% Pod 5 switch 1 host 2
(34.4363333333333,-0.95) circle (0.15);
\draw (35.1756666666666,0) -- (35.1756666666666,-0.8)		% Pod 5 switch 2 host 0
(35.1756666666666,-0.95) circle (0.15);
\draw (35.715,0) -- (35.715,-0.8)		% Pod 5 switch 2 host 1
(35.715,-0.95) circle (0.15);
\draw (36.2543333333333,0) -- (36.2543333333333,-0.8)		% Pod 5 switch 2 host 2
(36.2543333333333,-0.95) circle (0.15);
;

\end{tikzpicture}